# Contribution of magnetism to the Saturn rings origin

## Vladimir V. Tchernyi* and Sergey V. Kapranov †

*Modern Science Institute, SAIBR. Osennii blvd, 20-2-702, Moscow, 121614, Russia. Email: chernyv@bk.ru

†A.O. Kovalevsky Institute of Marine Biological Research, Russian Academy of Sciences. Nakhimova Ave. 2, Sevastopol 299011, Russia. Email: sergey.v.kapranov@yandex.ru

**Abstract**

The magnetization relationships for magnetically uniform spherical particles of the Saturn rings are derived. The problem of a solitary magnetized sphere and spherical particle among identical particles scattered in a disk-like structure is solved. The differential equations of motion of particles in the gravitational and magnetic field are derived. Special cases of these equations are solved exactly, and their solutions suggest that the superposition of the gravitational attraction and repulsion by a magnetic field of the iced particles which possess diamagnetism can account for the stability of Saturn rings.

1. Introduction

At present, there are a number of hypotheses on the genesis of planetary rings [1]. (1) The rings could have been generated by the particles seceded from moons of the outer planets by collision with comets or meteorites. (2) A moon of the planet could have been disrupted by a passing celestial body. (3) The ring particles can be debris of a large comet tidally broken by the planet. (4) The rings can be the relic of a protosatellite disk. (5) The particles can be continuously forming as a result of volcanic activity on a moon of the planet.

Several theories that have been developed to date to explain the evolution and stability of Saturn's rings [2]-[7] postulate that the orbits of the ring particles are close to the equatorial plane of the planet, but none of these approaches consistently explains this peculiarity and observed electromagnetic phenomena [8]. In [9]-[12], superconductivity of rings' particles was assumed responsible for the location of Saturn's rings in the plane of its magnetic equator and several observed electromagnetic phenomena were explained. However, to date there is no experimental evidence that space ice may have superconductivity.

Particles of Saturn's rings are known to consist mostly of ice [13], which is diamagnetic at not too high pressures [14]. Below the atmospheric pressure, there are two stable polymorphs: the high-temperature ice Ih and the low-temperature ice XI (Fig. 1) [15, 16]. Unlike the proton-disordered ice Ih, the proton-ordered ice XI is stable only below ~73 K. From an infrared image taken by the Cassini spacecraft, the temperature of Saturn's rings has been found to vary from 70 K to 110 K [17], and it is below 73 K over the entire area of the rings at equinox [18]. Thus, it is likely that the particles of Saturn's rings consist mostly of ice XI.

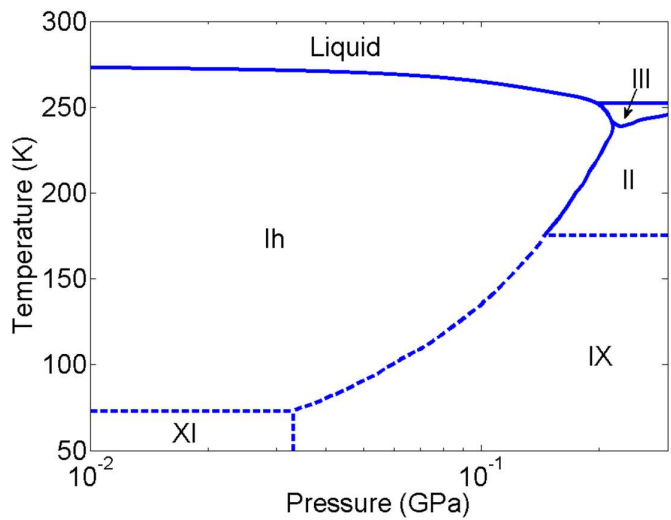

Fig. 1: Low-pressure part of phase diagram of ice. Adapted from [15].

Yet, it is shown in this paper that diamagnetic rings particles may be responsible for the rings origin and stability. All diamagnetic materials in non-uniform magnetic fields are known to be expelled into the weak-field areas, and this feature provides a minimum of the potential energy of diamagnetic particles in the magnetic equator plane. Taking this into account, the role of magnetism in the stabilization of planetary rings can be fairly universal. To understand the evolution of Saturn's rings and to shed light on their origin, magnetization and dynamics of ring particles in the magnetic field of the planet should be studied. For this purpose, equations of collisionless motion of diamagnetic particles in Saturn's magnetic field will be considered and particular cases of these equations will be solved to demonstrate their physical implications.

## 2. Magnetostatics of Uniformly Magnetized Spheres in Magnetic Field

### 2.1. Sole Uniformly Magnetized Sphere

The Maxwell-Ampère circuital law equation [19] reads

$$\nabla \times \mathbf{B} = \mu_0 \left( \mathbf{J} + \varepsilon_0 \frac{\partial \mathbf{E}}{\partial t} \right). \tag{1}$$

where $\mu_0$ and $\varepsilon_0$ are the vacuum permeability and permittivity, respectively. Both the current density and rate of electric field variation in the particle are zero ($\mathbf{J} = \mathbf{0}$ and $\partial \mathbf{E}/\partial t = \mathbf{0}$) because there is no macroscopic current in the particle and the electric field is assumed vanishingly weak. Then, one obtains

$$\nabla \times \mathbf{B} = \mathbf{0}. \tag{2}$$

If the curl of magnetic field is zero, there exists a scalar magnetic potential $\Phi$, which is the solution of the Laplace equation

$$\nabla^2 \Phi = 0. \tag{3}$$

In general, the solution of (3) in spherical coordinates is

$$\Phi(r,\theta,\varphi) = \sum_{l=0}^{\infty} \sum_{m=-l}^{l} \left( A_l r^l + C_l r^{-l-1} \right) P_l^m(\cos\theta) e^{-im\varphi} \tag{4}$$

where $P_l^m$ is the associated Legendre polynomial of degree $l$ and order $m$, and $A_l$ and $C_l$ are constants that are found from the boundary conditions [19].

The azimuthal symmetry of the magnetic field and potential implies that $m = 0$ and

$$\Phi(r,\theta) = \sum_{l=0}^{\infty} \left( A_l r^l + C_l r^{-l-1} \right) P_l(\cos\theta). \tag{5}$$

The magnetic potential (5) must be finite at $r = 0$ and at $r = \infty$, hence the radial functions for the potential inside ($r \leq R$) and outside ($r \geq R$) the sphere with radius $R$ must be different:

$$\Phi_{in}(r,\theta) = \sum_{l=0}^{\infty} A_l r^l P_l(\cos\theta), \quad r \leq R, \tag{6}$$

and

$$\Phi_{out}(r,\theta) = \sum_{l=0}^{\infty} C_l r^{-l-1} P_l(\cos\theta), \quad r \geq R. \tag{7}$$

Magnetic field intensity is found as the negative gradient of the magnetic potential:

$$\mathbf{H} = -\nabla\Phi. \tag{8}$$

Its radial component is discontinuous at $r = R$ due to the non-zero surface magnetic charge density $\sigma_M$ found from magnetization $\mathbf{M}$ according to Gauss's divergence theorem:

$$\sigma_M = \mathbf{n} \cdot \mathbf{M} = M\cos\theta. \tag{9}$$

Thus, the first boundary condition is

$$\left(\frac{\partial \Phi_{in}}{\partial r} - \frac{\partial \Phi_{out}}{\partial r}\right)\bigg|_{r=R} = \sigma_M. \tag{10}$$

From (6)-(10), one arrives at the equality

$$\sum_{l=0}^{\infty} l A_l R^{l-1} P_l(\cos\theta) + \sum_{l=0}^{\infty} (l+1) C_l R^{-l-2} P_l(\cos\theta) = M\cos\theta. \tag{11}$$

This equality is fulfilled only if the Legendre polynomials are of degree 1 ($l = 1$). Then,

$$A_1 + 2C_1 R^{-3} = M. \tag{12}$$

The second boundary condition results from the continuity of the magnetic potential at $r = R$:

$$(\Phi_{in} - \Phi_{out})\big|_{r=R} = 0 \tag{13}$$

or alternatively, from the continuity of the tangential components of the magnetic field intensity at the surface of the sphere with radius $R$:

$$\left(\frac{\partial \Phi_{in}}{\partial \theta} - \frac{\partial \Phi_{out}}{\partial \theta}\right)\bigg|_{r=R} = 0. \tag{14}$$

From the boundary condition (13) or (14) applied to the magnetic potentials (6) and (7) with $l = 1$,

$$A_1 = C_1 R^{-3}. \tag{15}$$

Solving the system of equations (12) and (15), one obtains the integration constants

$$A_1 = M/3 \tag{16}$$

and

$$C_1 = MR^3/3. \tag{17}$$

Because, by definition, magnetization is magnetic moment **m** per unit volume:

$$\mathbf{m} = (4/3)\pi R^3 \mathbf{M},$$

the magnetic potential in (6) and (7) can be written as

$$\Phi_{in}(r,\theta) = (Mr\cos\theta)/3 = \mathbf{m}\cdot\mathbf{r}/(4\pi R^3) \quad \text{for } r \leq R, \tag{18}$$

and

$$\Phi_{out}(r,\theta) = MR^3 \cos\theta/(3r^2) = \mathbf{m}\cdot\mathbf{r}/(4\pi r^3) \quad \text{for } r \geq R. \tag{19}$$

From (18) and (19), one can derive the magnetic field intensity inside the sphere:

$$\mathbf{H}_{in} = -\nabla\Phi_{in} = -M(\mathbf{e}_r \cos\theta - \mathbf{e}_\theta \sin\theta)/3 = -\mathbf{M}/3 \tag{20}$$

and outside the sphere:

$$\mathbf{H}_{out} = -\nabla\Phi_{out} = MR^3(2\mathbf{e}_r \cos\theta + \mathbf{e}_\theta \sin\theta)/(3r^3) = m(2\mathbf{e}_r \cos\theta + \mathbf{e}_\theta \sin\theta)/(4\pi r^3) \tag{21}$$

where $\mathbf{e}_r$ and $\mathbf{e}_\theta$ are the radial and polar unit vectors of the spherical coordinate system.

Now, consider a sphere placed in an external magnetic field $\mathbf{B}_0$ with intensity $\mathbf{H}_0$. Inside the sphere, the total magnetic field is the sum of the external magnetic field and the fields due to the induced current and magnetization:

$$\mathbf{B} = \mathbf{B}_0 + \mathbf{B}_{in} = \mathbf{B}_0 + \mu_0(\mathbf{H}_{in} + \mathbf{M}) = \mathbf{B}_0 + 2\mu_0\mathbf{M}/3 \quad \text{for } r \leq R, \tag{22}$$

in which (20) is taken into account. The corresponding magnetic field intensity is

$$\mathbf{H} = \mathbf{H}_0 + \mathbf{H}_{in} = \mathbf{B}_0/\mu_0 - \mathbf{M}/3 \quad \text{for } r \leq R. \tag{23}$$

The total magnetic field is the total magnetic field intensity multiplied by the absolute permeability $\mu$:

$$\mathbf{B} = \mu \mathbf{H} \tag{24}$$

Then, it follows from (22)-(24) that

$$\mathbf{B}_0 + 2\mu_0 \mathbf{M}/3 = \mu(\mathbf{B}_0/\mu_0 - \mathbf{M}/3) \quad \text{for } r \leq R, \tag{25}$$

whence

$$\mathbf{M} = \frac{3\mathbf{B}_0}{\mu_0} \frac{\mu - \mu_0}{\mu + 2\mu_0} \tag{26}$$

and

$$\mathbf{m} = \frac{4\pi R^3 \mathbf{B}_0}{\mu_0} \frac{\mu - \mu_0}{\mu + 2\mu_0}. \tag{27}$$

Equations (26) and (27) are, in fact, magnetic equivalents of the Clausius–Mossotti relation derived for polarization of dielectrics. As with this relation, they do not hold if the particle is magnetically heterogeneous on the macroscopic or microscopic scale, e.g. it contains spatially separated magnetic dipoles.

As it follows from (26) and (27), magnetization and magnetic moment of diamagnetic spheres are opposing to the external magnetic field. Among diamagnetics, superconductors have the highest absolute values of magnetization ($-1.5\mathbf{B}_0/\mu_0$) and magnetic moment ($-2\pi R^3 \mathbf{B}_0/\mu_0$) and the lowest permeability ($\mu = 0$ N A$^{-2}$).

### 2.2. Uniformly Magnetized Sphere in a Disk-Shape Structure Consisting of the Same Spheres

Consider a sphere placed in magnetic field among the same magnetized spheres in an infinite disk structure. Such a disk is a model of dense rings of Saturn. The spheres are assumed to be uniformly spread in hexagonal packing with the planar density $\sigma$.

The magnetized spheres, each having the magnetic dipole moment **m**, will then collectively generate the induced global magnetic field, which is convenient to use as an integral characteristics instead of than the summed magnetic field intensities of discrete spheres. If the angle between the disk and magnetic equator plane is $\vartheta_t$ and the distance from the center of the sphere to the nearest edge of the disk representing the centers of the nearest particles is $r_0$, a magnetized sphere will experience the total magnetic field intensity $\mathbf{H}_d$ of the disk found from (21):

$$\mathbf{H}_d = \int_S \mathbf{H}_{out} \sigma ds = \frac{\sigma m}{4\pi} \int_0^{2\pi} d\varphi \int_{r_0}^{\infty} \sin\varphi \cos\vartheta_t \frac{2\mathbf{e}_r \sqrt{1-\sin^2\varphi\cos^2\vartheta_t} + \mathbf{e}_\theta \sin\varphi\cos\vartheta_t}{r^2} dr \quad (28)$$
$$= -\mathbf{e}_\theta \sigma m \cos^2\vartheta_t / (4 r_0).$$

In (28), the integration is performed throughout the disk area, and the disk is assumed continuous. Drawing an analogy with (22) and (23), the total magnetic field inside the sphere is

$$\mathbf{B} = \mathbf{B}_0 + \mathbf{B}_{in} + \mathbf{B}_d = \mathbf{B}_0 + \mu_0 (\mathbf{H}_{in} + \mathbf{M} + \mathbf{H}_d) = \mathbf{B}_0 + \mu_0 \mathbf{M}(2 - \pi R^3 \sigma \cos^2\vartheta_t / r_0)/3 \quad \text{for } r \leq R, (29)$$

and its intensity is

$$\mathbf{H} = \mathbf{H}_0 + \mathbf{H}_{in} + \mathbf{H}_d = \mathbf{B}_0/\mu_0 - \mathbf{M}(1 - \pi R^3 \sigma \cos^2\vartheta_t / r_0)/3 \quad \text{for } r \leq R. \quad (30)$$

After the similar algebra as in the derivation of (26) and (27), the magnetization is

$$\mathbf{M} = \frac{3\mathbf{B}_0}{\mu_0} \frac{\mu - \mu_0}{\mu + 2\mu_0 - (\mu + \mu_0)\pi R^3 \sigma \cos^2\vartheta_t / r_0}, \quad (31)$$

and the magnetic moment is

$$\mathbf{m} = \frac{4\pi R^3 \mathbf{B}_0}{\mu_0} \frac{\mu - \mu_0}{\mu + 2\mu_0 - (\mu + \mu_0)\pi R^3 \sigma \cos^2\vartheta_t / r_0}. \quad (32)$$

The magnetization of superconductors in this case is $-3\mathbf{B}_0/(2\mu_0 - \mu_0 \pi R^3 \sigma \cos^2\vartheta_t / r_0)$ and their magnetic moment is $-4\pi R^3 \mathbf{B}_0/(2\mu_0 - \mu_0 \pi R^3 \sigma \cos^2\vartheta_t / r_0)$. It is easy to see that the magnetization and magnetic moment are higher in their absolute values than those of a sole sphere. Hence, the force of diamagnetic expulsion into the weak-field areas in this case is expected to be stronger.

### 3. Potential Energy and Equations of Motion of Magnetized Spheres

The magnetic moments in (27) and (32) can be generalized by the formula

$$\mathbf{m} = C\mathbf{B}_0 \tag{33}$$

where $C$ is a function of magnetic properties and size of the spheres. For diamagnetics, $C < 0$.

For the derivation of the equations of motion, the potential energy $U$ of a magnetized sphere should be found. For a sphere with mass $M$ in the magnetic and gravitational field, it is

$$U = -GM_S M/r - \mathbf{m} \cdot \mathbf{B}_{out} = -GM_S M/r - C\mu_0^2 m^2 \left(3\cos^2\theta + 1\right) \big/ \left(4\pi r^3\right)^2 \tag{34}$$

where $G = 6.67408 \cdot 10^{-11}$ m³·kg⁻¹·s⁻² is the gravitational constant and $M_S = 5.68319 \cdot 10^{26}$ kg is the mass of the planet (Saturn). The first term in the right-hand side of (34) is the gravitational component and the second one is the magnetic potential energy.

The force acting on the sphere is equal to the negative gradient of the potential energy:

$$M\mathbf{a} = -\nabla U = -\frac{\partial U}{\partial r}\mathbf{e}_r - \frac{1}{r}\frac{\partial U}{\partial \theta}\mathbf{e}_\theta = \mathbf{e}_r\left[-\frac{GM_S M}{r^2} - \frac{3C\mu_0^2 m^2}{8\pi^2 r^7}\left(3\cos^2\theta + 1\right)\right] - \mathbf{e}_\theta \frac{3C\mu_0^2 m^2}{8\pi^2 r^7}\sin\theta\cos\theta \tag{35}$$

where $\mathbf{a}$ is the linear acceleration.

From the formula of the acceleration vector in spherical coordinates [20], the respective equations of motion are

$$\begin{cases} \ddot{r} - r\dot{\theta}^2 - r\dot{\varphi}^2 \sin^2\theta = -(\nabla U)_r / M = -\dfrac{GM_S}{r^2} - \dfrac{3C\mu_0^2 m^2}{8\pi^2 r^7 M}\left(3\cos^2\theta + 1\right), \\ r\ddot{\theta} + 2\dot{r}\dot{\theta} - r\dot{\varphi}^2 \sin\theta\cos\theta = -(\nabla U)_\theta / M = -\dfrac{3C\mu_0^2 m^2}{8\pi^2 r^7 M}\sin\theta\cos\theta, \\ r\ddot{\varphi}\sin\theta + 2\dot{r}\dot{\varphi}\sin\theta + 2r\dot{\theta}\dot{\varphi}\cos\theta = -(\nabla U)_\varphi / M = 0. \end{cases} \tag{36}$$

Equations (36) are analytically unsolvable. The long-term (on a cosmic time scale) numerical solution of equations (36) is neither feasible nor informative because integration over thousands of oscillations with only one set of initial conditions can be performed at one run. For this reason, two most important special cases of the equations of motion giving meaningful results will be considered below.

### 3.1. Particles Orbiting at Equal Distance from the Centre ($r$ = const) and Experiencing Zero Magnetic Force ($C = 0$)

This is the classical circular motion around the centre of mass of the planet. Because in this case the radial acceleration ($\ddot{r}$) and radial velocity ($\dot{r}$) are zero, the equations of motion (36) are reduced to a redundant system

$$\begin{cases} \dot{\theta}^2 + \dot{\varphi}^2 \sin^2\theta = GM_S/r^3, \\ \ddot{\theta} - \dot{\varphi}^2 \sin\theta \cos\theta = 0, \\ \ddot{\varphi} + 2\dot{\theta}\dot{\varphi}/\tan\theta = 0. \end{cases} \quad (37)$$

From the first and second equations in (37),

$$\ddot{\theta} + \dot{\theta}^2 \cot\theta = GM_S \cot\theta/r^3, \quad (38)$$

whose solution is

$$\theta = \arccos\left[-\dot{\theta}_0 \sqrt{r^3/(GM_S)} \sin\left(t\sqrt{GM_S/r^3}\right)\right], \quad (39)$$

where the initial angle without loss of generality is taken to be $\theta_0 = \pi/2$ (at $t = 0$). The azimuthal velocity $\dot{\varphi}$ is then determined as

$$\dot{\varphi} = \frac{GM_S \sqrt{GM_S/r^3 - \dot{\theta}_0^2}}{GM_S - r^3 \dot{\theta}_0^2 \sin^2\left(t\sqrt{GM_S/r^3}\right)}. \quad (40)$$

One should seek the initial velocity $\dot{\theta}_0$ from the thickness $b$ of main rings of Saturn ($b \approx 10$ m). Expanding (39) in the Taylor series near $\pi/2$, one can obtain

$$\theta \approx \pi/2 + \dot{\theta}_0 \sqrt{r^3/(GM_S)} \sin\left(t\sqrt{GM_S/r^3}\right), \quad (41)$$

whence the difference $\Delta\theta$ between the maximum and minimum zenith angles is

$$\Delta\theta \approx 2\dot{\theta}_0 \sqrt{r^3/(GM_S)} = b/r, \quad (42)$$

and the initial zenithal velocity is

$$\dot{\theta}_0 = b\sqrt{GM_S/r^5}\big/2. \qquad (43)$$

From (43), the initial angular velocity at the mean ring radius of 105,000 km is 1.7·10$^{-12}$ s$^{-1}$, which gives the linear velocity 0.2 mm·s$^{-1}$. This value is at odds with the Keplerian orbital velocity at the same radial distance: $r\dot{\varphi} \approx \sqrt{GM_S/r} = 19$ km·s$^{-1}$. It is highly unlikely that the velocity components of colliding particles of the rings are so different.

On the other hand, if the velocity components were comparable, i.e. $\dot{\theta}_0 = \sqrt{GM_S/r^3}$, one would arrive at $\theta = \arccos\left[-\sin\left(t\sqrt{GM_S/r^3}\right)\right]$, a triangular periodic function of time with the extreme values of 0 and $\pi$ (i.e. the particles would wobble between the planetary poles), which is in complete disagreement with the flat structure of the rings.

### 3.2. Magnetized Particles Orbiting at Equal Distance from the Centre ($r$ = const)

This is the case of $C \neq 0$, and the zenith angle equation is

$$\ddot{\theta} + \dot{\theta}^2 \cot\theta = \frac{GM_S}{r^3}\cot\theta + \frac{3C\mu_0^2 m^2}{2\pi^2 r^8 M}\cot\theta \cos^2\theta. \qquad (44)$$

The MuPAD symbolic calculator of Matlab environment gives the following singular solution of (44):

$$\theta = \pi/2. \qquad (45)$$

The azimuthal velocity in this case is

$$\dot{\varphi} = \sqrt{GM_S/r^3 + 3C\mu_0^2 m^2\big/\left(8\pi^2 r^8 M\right)}, \qquad (46)$$

which implies that the gravitational force in the stable circular orbiting is counterbalanced both by the centrifugal force and the force of diamagnetic expulsion.

The singularity of the solution (45) accounts for the fact of essentially planar structure of Saturn's rings which lie in the magnetic equator plane.

### 4. Conclusions

In this work, magnetostatics and dynamics of magnetized spheres in the spherically symmetric gravitational and axially symmetric planetary magnetic field has been presented. Relationships for the magnetization and magnetic moment of both a sole sphere and a sphere placed in an infinite disk-like structure of evenly distributed identical spheres modeling Saturn's dense rings have been found. The magnetic force acting on a sphere in the "disk" with magnetized spheres is stronger than that acting on a single sphere.

The potential energy relationship for the spherical particles in the superposition of the gravitational and magnetic fields has been derived. It has resulted in formulating equations of collisionless motion of the ring particles. Two special cases of these equations have been considered: with the particle in zero and non-zero magnetic field at a constant distance from the gravitation centre. In the gravitational field only (i.e., if the magnetic field is zero), the ratio of the particle's angular velocity components has proven to be extremely unlikely, which apparently disproves the purely gravitational theory of stability of Saturn's rings. If the additional axially-symmetric magnetic force is exerted on the particles, their circular orbits fall on the magnetic equator plane, as it has followed from the equation solution and as established in several spacecraft missions to the outer planets.